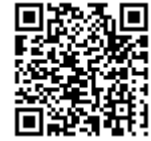

Research Article

# Innovating HR Using an Expert System for Recruiting IT Specialists – ESRIT

**Ciprian-Octavian Truică and Adriana Barnoschi**

Faculty of Automatic Control and Computers, University "Politehnica" of Bucharest, Romania

Correspondence should be addressed to: Ciprian-Octavian Truică; ciprian.truica@cs.pub.ro

**Abstract**

One of the most rapidly evolving and dynamic business sector is the IT domain, where there is a problem finding experienced, skilled and qualified employees. Specialists are essential for developing and implementing new ideas into products. Human resources (HR) department plays a major role in the recruitment of qualified employees by assessing their skills, using different HR metrics, and selecting the best candidates for a specific job. Most recruiters are not qualified to evaluate IT specialists. In order to decrease the gap between the HR department and IT specialists, we designed, implemented and tested an Expert System for Recruiting IT specialist – ESRIT. The expert system uses text mining, natural language processing and classification algorithms to extract relevant information from resumes by using a knowledge base that stores the relevant key skills and phrases. The recruiter is looking for the same abilities and certificates, trying to place the best applicant into a specific position. The article presents a developing picture of the top major IT skills that will be required in 2014 and it argues for the choice of the IT abilities domain.

**Introduction**

Many organizations have gone beyond the traditional functions and developed human resource management information systems, which support recruitment, selection, hiring, job placement, performance appraisals, employee benefit analysis, health, safety and security.

A Human Resources Management System (**HRMS**) or Human Resources Information System (**HRIS**) refers to the systems and processes at the intersection between human resource management (**HRM**) and information technology. [1]

Human resources (HR) departments have a generally and common administrative function in all organizations. The HR function consists of the recording about employee data that usually includes personal histories, skills, capabilities, accomplishments and salary. Organizations could formalize the processes of selection, evaluation, and payroll. In order to reduce the manual work from these activities, most companies began to automate these processes by introducing specialized human resource management systems.

The system presented in the article (called as expert system for recruiting IT

___

specialists - ESRIT) provides the ability to "read" the applications (resumes), enters relevant data into a knowledge base, and offers the placement advice (by individual calculated score of applicant) to the recruiter or employer. The system is designed to work in the domain of IT professionals and project managers (or leadership) for IT projects and services. The choice was made considering the following:

1) The developing picture of the top major **IT skills that will be in requirement in 2014** [2] represented in the hiring and salary surveys, such as the *2014 IT Skills and Salary Survey* from Global Knowledge and Windows IP Pro, TEKsystems' 2014 Annual IT Forecast, Foote Research Group's 2014 IT Skills and Certifications Pay Index, Computerworld's annual Forecast survey, Robert Half Technology Survey, and information from the U.S. Bureau of Labor Statistics, Futurestep, Mondo, GovLoop, and Dice.

2) Though 2013**, project management** continues to be a highly desired skill. Some surveys place it right behind software developers/engineers, with demand that has grown more than 10 percent from last year. Some of the latest survey results show that the demand for project managers is resurging in more complex and strategic long-term technology plans.

3) **Social business** represents a significant transformational opportunity for any organization. The article presents a new IBM Institute for Business Value study, based on responses from more than 1,100 individuals and interviews with more than two dozen executives from leading organizations, reveals how organizations can use social approaches to create meaningful business value. The paragraph also examines the role social media could play in presenting a strategic view of customer data and looks at the implications for IT specialists.

4) The business rewards of participating in social media have become too great to ignore for enterprises in regulated industries. Social media presents unique **compliance** challenges. It is imperative to use a strategy for monitoring and engaging customers on social media in a compliant fashion.

5) Panic over losing control is one natural barrier into implementing social media across an organization. Another unique ***security feature*** is limiting permissions that set control over publishing firmly in the hands of those who are most trusted. When it comes to delivering secure, new business services, identity is the crucial element that bridges the gap between growing and protecting the business.

6) An important directive of any IT department (a key element in business success) is keeping applications running without difficulties. Well-performing applications help internal users work efficiently and productively, with decreasing costs. While most companies use some level of **APM *(Application Performance Management),*** they find it challenging to keep up with increasingly complex IT environments and business demands, according to a recent survey conducted by IDG Research Services for CA Technologies.[3]

In most cases, IT is judged based on the success or failure of the projects; so, organizations decide to make investments in business analyst/project manager skills. For the success of the project to be more visible, project managers must gather diverse business and end-user requirements, set priorities, must be able to talk with developers about the technologies involved. Also, they must ensure that the processes are completed in a timely manner and tested methodically before they are rolled out.

Recruiting is the main method used by HR departments to acquire potential candidates for available positions within an organization. **ESRIT** (**E**xpert **S**ystem for

---

**R**ecruiting **IT** Specialists) performs the function of human resource management that involves the recruitment and the placement of the applicants (next possible employees) of an organization. The system is designed to work in the domain of IT professionals and project managers (or leadership) for IT projects and services.

**IT Skills – Insight View**

UK IT and digital companies are among the worst affected by a skills shortage, which is forcing employers to look abroad for talent, according to City & Guilds. Research by the vocational training organization revealed that 60% of all employers are struggling to find the right candidates. The survey rises to 74% in the digital, IT and information services industry.

The Office of Fair Trading (OFT) is launching a formal investigation into the market for supplying IT and communications to the public sector. The OFT's examination of "whether competition in this sector could work better and the reasons why it may not be working as well as it could" follows a call to suppliers and buyers of IT services to the public sector to provide information about their experiences.

UK businesses call for support to close virtual security skills gap. A quarter of UK organizations do not have the knowledge to manage virtual security deployments. Over half of these say this is due to a lack of training or funds available to train, according to a study by security firm Trend Micro. At an industry level, 57% of UK businesses want to see virtual security guidelines put in place to help organizations understand the best practice.

Costas Markides, professor of strategy and entrepreneurship at the London Business School, predicts that technology will make organizations less hierarchical, more decentralized and more democratic – in other words more human. "The implication is becoming very clear. We are moving away from running organizations like a machine to running them like a brain. We are moving from rules based world to a purpose-based world," he says.

The trend will have radical implications on the way organizations manage their employees. IT departments in particular, should give up trying to control and restrict how employees use technology and should instead refocus their energies on developing a sense of corporate purpose.

The survey "2014 Priorities Europe" (in which 85% of respondents work in corporate IT over Europe, 11% of respondents represents business analysts who work with IT departments) says:

- 80% will implement in 2014 a formal governance, risk and compliance program;

- 25% of their IT organizations still stuck in recession;

- 29% of their IT departments expend IT to support business growth and 20% help the business automate more;

- 25% of expenses represent the costs for implementing business applications in HR management.

**Social Media for Business Purposes in Compliant Manner**

With clients and employees dispersed across the world, organizations needed to find a cost efficient and effective way to *communicate* with team members and customers, to share documents and screens, and to invite guests to join video conferencing meetings.

Social business is defined as embedding social tools, media, and practices into the ongoing activities of the organization. Social business enables individuals to connect and share information and insights more effectively with others, both inside and outside the organization. [4]

IT can help guide the integration of social media strategy and technologies into the wider enterprise. IT has an opportunity to influence both the short-term needs and

___________

the long-term requirements and integrate social media applications with the operational systems that provide reporting and analysis of business performance.

IT can help in the selection of social media tools to help groups like market research be confident that the data they view is accurate and validated. IT can ensure smooth integration between unstructured and structured pools of data for analysis, such as unstructured social media data and structured data from a company's transactional databases.

Ray Wang of Constellation Research makes the point that, as guardians of data and its distribution within the enterprise, IT has ample opportunity to be engaged in social media strategy.[5]

Social media networks are re-shaping the way organizations engage their customers and nurture their relationship to brands, products and services. Here are some figures that give an idea of the scale of the social media phenomenon:

- 1.43 billion people worldwide visited a social networking site last year.

- Nearly 1 in 8 people worldwide have their own Facebook® page.

- Last year, one million new accounts were added to Twitter® every day.

- Three million new blogs come online every month.

- 65 percent of social media users say they use it to learn more about brands, products and services.

Legal requirements are very different across industries and international borders (by line of business, geography or brand), but the principles of social media compliance are the same for every enterprise. To ensure that policies are consistently enforced, an organization needs effective review procedures, reliable records retention, and global awareness of social media activity. All of these elements are attainable when the right technologies are put in the hands of administrators, compliance officers and educated employees.

Companies must comply with regulations such as Know Your Customer (KYC), export/import laws, PCI/DSS data security standards, and taxation policies. They must also comply with the SLAs that they have committed to their customers. If they don't meet these requirements they can face serious fines, penalties and loss of business.

Business processes that deliver operational efficiency, business visibility, and agility give an enterprise an edge by enabling it to conduct business in a cost-effective, dynamic way. As organizations strive for greater efficiency and effectiveness, they create or adapt technology to fill their needs. BPM, Social, and Mobile technologies are helping to drive a fundamental business transformation. The organizations who are implementing these technologies respond to today's multi-faceted business challenges and take advantage of new opportunities.[6]

When it comes to delivering secure, new business services, identity is the crucial element that bridges the gap between growing and protecting the business.

**ESRIT – Expert System for Recruitment IT Specialists**

The European Union's free movement of people opened for the work market new opportunities. Now, every European citizen can live and work in any of the countries that have signed and ratified the treaty. A great number of resumes, from all over Europe, flood the recruitment and job placement agencies. This volume of resumes makes it difficult for the employees of these agencies to search for the best candidate for a particular job. Another problem that arises consists of the resumes that have various formats, Word documents (different versions), Adobe PDFs, and so on, and this issue makes the search very difficult.

Given a set of resumes and a set of job descriptions the system must determine the resumes that best match for a specific job. This is a classification problem that can be solved with a binary classifier. If a resume best matches with the job requirements it will be placed into the first class, otherwise it will be tagged with *not applicable* (it means it will be put in the second class). The classification attributes are the key elements in the job description. The best approach to determine the classification attributes is extracting the key skill from the job description. Then, the system tries to find the attributes in the resumes, by giving a fitting score to each resume, which will be called the *Resume Score* - $R_s$ (formula 1). The key skills will be divided into two categories: the first category consists of a set of key word, which best describes the required technologies; the second category is a set of key phrases, which represents the best skills required for the job. These categories, including the name of the job, will populate a knowledge base. The knowledge base stores the information as dictionaries – a collection of keys and values; each job will have two dictionaries, one for words and one for phrases. Each dictionary will keep in the key field the word or phrase and in the value field the importance score ($s(w_i)$ or $s(p_i)$) given by the user. ESRIT stores this information into a NoSQL DBMS, because the database query response is faster that the query response of a RDBMS. Due to the fact that the knowledge base is a collection of documents, the best candidate for the NoSQL DBMS was MongoDB.

Regarding the resumes, the system first uses natural language processing to lemmatize the words, so that the base of each word is extracted, considerably minimizing the number of words in the text. The second step of resumes processing uses text mining techniques together with the sets of words and phrases from the knowledge base in order to calculate the resume scores. Thereby, the system correctly classifies the resumes, and determines if a resume matches the job description. The resume score is also used by the human recruiter to take the decision if the resume needs their attention or not. The system uses the NLTK python package for processing the resume text.

The appearance coefficient (ac) of a word in a resume is calculated as the number of occurrences of that word in the resume divided by the total number of words (formula 2). The appearance coefficient of a phrase is calculated as the number of occurrences of the phrase and all the resulting phrases from the different permutations of the words from the original phrase, divided by the total number of phrases – with the same number of words as the searched phrase – which appear in the resume text (formula 3).

For a resume, the score is calculated using the following formula:

$$R_s = \sum_{i=1}^{n} ac(w_i) \cdot s(w_i) + \sum_{i=1}^{m} ac(p_i) \cdot s(p_i) \ (1)$$

$$ac(w_i) = \frac{count(w_i)}{count(w)} \ (2)$$

$$ac(p_i) = \frac{count(p_i)}{count(p)} \ (3)$$

Where:

- $R_s$ is the resume score

- $n$ is the number of searched words

- $m$ is the number of searched phrases

- $ac(w_i)$ is the appearance coefficient of the word $w_i$ in the resume

- $s(w_i)$ is the importance score of word $w_i$, this score is extracted from the knowledge base

- $ac(p_i)$ is the appearance coefficient of phrase $p_i$ in the resume

- $s(p_i)$ is the importance score of phrase $p_i$, this score is extracted from the knowledge base

- $count(w_i)$ is the number of occurrences of the word $w_i$ in the resume

- $count(w)$ is the total number of words

- $count(p_i)$ is the number of occurrences of the phrase $p_i$ and its permutations in the resume

- $count(p)$ is the total number of phrases that has the length of phrase $p_i$

**Design and Implementation**

The ESRIT is implemented using Python and the MongoDB DBMS to store the knowledge base. The database is used to store the dictionaries of words and phrases with their respective scores.

The first step: the system parses and gets all the words from each resume. This module will accept a list of documents and then return a dictionary where the key field is the resume *id* and the value field is a string of characters. The string of characters is created by using the Linux abiword terminal tool that receives three parameters: the first is *–t* that tells the program to do the parsing, the second parameter is the name of document in the new format and the third parameter is the document in the old format. Using a regular expression, during this step, any punctuation sign and special characters will be removed. The algorithm for this module is:

getDocumentText(listDocuments):

foreach document in listDocuments:

exec "abiword –t newdocument.txt document.name"

file = open(newdocument)

text = file.readlines()

file.close()

text = text.replace('\n', ' ')

text = regular_expresion.substitute('[^A-Za-z0-9]', ' ')

text= ' '.join(text.split())

dict[document.id] = text

return dict

The *getDocumentText* will return a dictionary of resume *id*-s and the text stripped of all punctuation and special characters.

The second module uses a lemmatization algorithm that parses all the words that have been extracted in the first step and gets the root word depending on the word location and its syntax in the text. Also, the lemmatization algorithm drops the English language stop words, e.g. "the", "an", "and", etc. The *lemmatizeText* function of the system will return the lemmatized text given as parameter.

For each resume, the next module *wordAppearanceCoefficient* counts the number of appearances of each word by calculating the appearance coefficient - $ac(w_i)$ – shown in formula 2. The module will return a dictionary with the next form {word1: oc1, word2: oc2, …}.

The fourth module is *wordScore* that searches for the specific words – taken from the knowledge base – in the resume and it calculates the words score – appearing in formula 1 as $\sum_{i=1}^{n} ac(w_i) \cdot s(w_i)$ – in order to compute the resume score. This module receives two parameters: the first one is the dictionary that was returned by the *wordAppearanceCoefficient* module, and the second parameter is the set of words with scores taken from the knowledge base. Each word from the knowledge base will be lemmatized because the algorithm searches for the word lemma.

The next module *phraseAppearanceCoefficient* counts the appearance coefficient for a search phrase, by giving three parameters: the first one is the resume lemmatized text, the second one is the lemmatized search phrase and the last one is the number of phrases, with a equal number of words as the number of words in the searched phrase, that appear in the resume.

The last module is the *phraseScore* that will search phrases, taken from the knowledge base, in the resume text and calculate the phrase score, appearing in formula 1 as $\sum_{i=1}^{n} ac(p_i) \cdot s(p_i)$. This module requires two parameters: the first one is the resume lemmatized text and the second one is a list of phrases with scores. Each phrase word will be lemmatized and then, during execution, new phrases will be created by permuting the words. Finally, there will be *n!* search phrases, where *n* is the number of words in the original search phrase. The total number of phrases – with the same number of words as the original searched phrase – is calculated during this step. This module will return the phrase score.

The *resume score* $R_s$ (formula 1) is given by summing the results of *wordScore* and *phraseScore* functions.

**ESRIT Knowledge Base**

ESRIT uses a database from which the system extracts the information to calculate the resumes' scores. The information is stored in a MongoDB database. For each KB job there is a record structured in two fields: the first contains the key words with their scores and the second consists of the key phrases with their scores. The decision to use MongoDB was taken due to the knowledge base structure that resembles with a collection of documents, and because it was the best candidate for storing, manipulating and processing this type of data format.

The knowledge base is built by the user and it will include the key words and phrases that appear in the job description. To build a record for a job, the user has to look at the job description and extract the key skills required for that job. After all skills were found, they are split into two different sets: words set and phrases set. The words set represent the mandatory and the nice to have technology names and certification abbreviations. The phrase set contains the required skills for the job. The scores will be given by the recruiter knowing which skills are mandatory and which ones are nice to have.

For example, for the Java programmer job, the record in the knowledge base will look something like:

```
{
"_id":
    zz ObjectId("5310d7c98528d412c33d4578"),

"name": "Java Programmer",

"words": [

    {"word": "Java ", "score": 10},

    {"word": "JSF ",  "score": 10},

    {"word": "J2EE ", "score": 10},

    {"word": "C/C++ ", "score": 5},

    {"word": "SQL ", "score": 8}
        ],
"phrases": [

{"phrase": "Oracle Database","score" : 7.5},

{"phrase": "professional java programmer",
    "score": 9},
        ]
    }
```

**Accuracy and Performance Testing**

ESRIT accuracy test was done in the following way: four job descriptions were selected from different recruitment sites. The selected jobs are: Java Programmer, Virtualization Engineer, Network Engineer and IT Project Manager. For each of these jobs, the knowledge base was built using the key words and the key phrases extracted from the job descriptions. A score was assigned to the most relevant words and phrases for a specific job. A recruiter selected and classified, for each test, 10 random resumes and he determined which resume best fits each job. The recruiter made a hierarchical top for the resumes, and he tagged as *not applicable* (N/A) the resumes that did not fit the job description. After that, the same 10 resumes were given as input data to the system. Table 1 presents the comparison results.

**Table 1: System – Human Accuracy Comparison**

| Java Programmer | | | Virtualization Specialist | | | Network Engineer | | | IT Project Manager | | |
|---|---|---|---|---|---|---|---|---|---|---|---|
| Resume | System | Human | Resume | System | Human | Resume | System | Human | Resume | System | Human |
| resume1 | 0.18018 | 2 | resume1 | 0.16456 | 2 | resume1 | 0.0 | N/A | resume1 | 0.03967 | 5 |
| resume2 | 0.08823 | 5 | resume2 | 0.02673 | 8 | resume2 | 0.0 | N/A | resume2 | 0.0204 | 6 |
| resume3 | 0.15663 | 3 | resume3 | 0.08558 | 6 | resume3 | 0.0 | N/A | resume3 | 0.00596 | 8 |
| resume4 | 0.13281 | 4 | resume4 | 0.14355 | 3 | resume4 | 0.10253 | 3 | resume4 | 0.10254 | 4 |
| resume5 | 0.24402 | 1 | resume5 | 0.0 | N/A | resume5 | 0.0 | N/A | resume5 | 0.01625 | 7 |
| resume6 | 0.0 | N/A | resume6 | 0.08338 | 7 | resume6 | 0.11461 | 2 | resume6 | 0.0 | N/A |
| resume7 | 0.08 | 6 | resume7 | 0.178 | 1 | resume7 | 0.15866 | 1 | resume7 | 0.0 | N/A |
| resume8 | 0.01497 | 8 | resume8 | 0.01497 | 9 | resume8 | 0.0 | N/A | resume8 | 0.75698 | 1 |
| resume9 | 0.0 | N/A | resume9 | 0.08813 | 5 | resume9 | 0.08902 | 4 | resume9 | 0.1913 | 3 |
| resume10 | 0.06878 | 7 | resume10 | 0.11904 | 4 | resume10 | 0.0 | N/A | resume10 | 0.32337 | 2 |

As seen in the tables, the best three chosen resumes selected by the system are the same as the ones selected by the recruiter.

System performance time was tested by measuring fifty response times, for five different sets of input data. The knowledge base for the performance test contained four jobs. The tests were done for ten, twenty, thirty, forty and fifty resumes. Figure no.1 shows the test results and the performance graphic.

**Conclusions**

The future of business engagement on all levels, whether it is from a sales, HR, or IT service, is essentially social. Any organization must be able to manage and enforce custom social media policies across different teams and social network accounts.

More than half of the surveyed HR managers report they have open positions that may stay that way, as qualified candidates are rare. Nearly half of them say they expect things to remain this way through the first quarter of 2014 or perhaps longer.

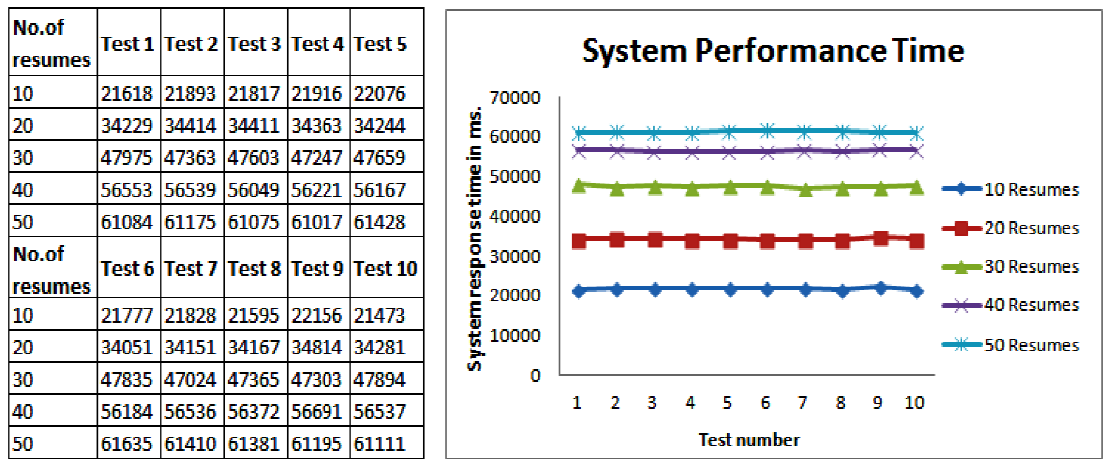

**Figure 1: System Performance Time**

Instead of waiting for a perfect applicant, about half of the employers surveyed are planning to train those who are short on the IT skills but have experience in the organization's field, up to 10 percent from 2013. About 25 percent are sending employees back to school. More than 25 percent of hiring managers plan to present their organizations to high school students or even younger. The plan is to begin luring future applicants to help them prepare. The solution is involving the employees into the organization's corporate goals. If they feel passionate about what the organization is trying to do, they will devote their energy to the company's goals rather than their own.

There is a growing trend: the use of new applications (software-driven business) by employees and clients. Customers are demanding new applications and a different experience, and employees require new tools to succeed and be productive. IT departments must evolve from single-source providers to business consultants; otherwise they risk to become marginalized in the technology-driven economy. This trend can also be seen in increased costs outside IT department.

In competitions for a job, very good candidates enroll in all EU Member States. There are hundreds of applicants per place. Recruiters must scan resumes making their hierarchy, placing on top 5 positions the competent candidates whose profiles made the best impression. The program highlights the achievements of the applicant, according to the job requirements. The system helps the recruiter to identify the applicant skills and it gets his attention from the first minutes by resume content. Hiring the wrong person for a specific job leads to high costs for the organization.

How often does the HR department measure its effectiveness? HR measurements are influential in the presentation of the areas where we could improve and better meet the needs of the organization and its employees. HR metrics provide meaningful data that help us make good decisions for the business and department.

On the other side of the barricade, for the job seekers, finding a job is not an exact science. The applicant must avoid mistakes into resume (the competition is fierce), he must highlight anything that makes him remarkable. He could not apply for the sake of applying. Mistakes happen; the applicant can be rejected based on subjective reasons. The idea is to move from HR metrics to HR analytics. HR analytics means the process of combining data mining with business analytics techniques in order to analyze human resources data. The goal of human resources analytics is to efficiently manage employees, so that business goals can be reached quickly and professionally.

___________

All of this activity is an encouraging sign for the economy as a whole, as it indicates that there is a continuing place for those who are willing to reinforce their business skills by adding IT skills to their resume or vice versa.

The system mean execution time is almost linear; processing and calculating the resume score, to determine if a resume is right for a job, is done in approximately second and minutes. For a large data volume, the system performs rather well. The mean execution times can be seen in Figure no. 2.

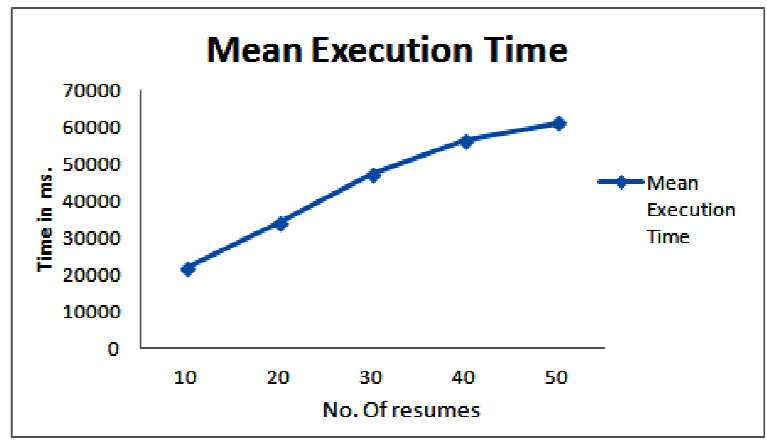

| No.of resumes | Mean Execution Time |
|---|---|
| 10 | 21814,9 |
| 20 | 34312,5 |
| 30 | 47526,8 |
| 40 | 56384,9 |
| 50 | 61251,1 |

**Figure 2: Mean Execution Time**

**Endnotes**

1. http://en.wikipedia.org/wiki/Human_resource_management_system

2. http://www.globalknowledge.com/training/generic.asp?pageid=3635

3. CA Technologies provides IT management solutions that help customers manage and secure complex IT environments to support agile business services. www.ca.com/apm

4. IBM Global Business Services

5. Ray Wang, Best Practices: Applying Social Business Challenges to Social Business Maturity Models, Constellation Research, 2011

6. Oracle WebCenter also helps organizations deliver contextual and targeted Web experiences to users and enables employees to access information and applications through intuitive portals, composite applications, and mash-ups. www.oracle.com/bpm

**References**

1. Bateman, K. (2014). "IT Budgets Expand as Downturn Recedes," *Computer Weekly's Digital Magazine for European IT Leaders* [online], [Retrieved February 2014], http://www.computerweekly.com/search/query?start=0&filter=1&q=IT+budgets+expand+as+downturn+recedes

2. Bird, S., Klein, E. & Loper, E. (2009). Natural Language Processing with Python, *O'Reilly Media Inc*.

3. CA Technologies, Market Pulse. "When It Comes to Application Performance, the User is Still King," http://www.ca.com/~/media/Files/whitepapers/when-it-comes-to-application-performance-the-user-is-still-king_eng.pdf

4. Goodwin, B. (2013). "Technology Will Make Business Less Hierarchical and More Human," *Computer Weekly's Digital*

___________